\documentclass{amsart}
\usepackage{amsaddr}

\usepackage{times}
\usepackage{yfonts}

\usepackage[utf8x]{inputenc}
\usepackage{amssymb}
\usepackage{amsmath}
\usepackage{graphicx} % more modern
\usepackage{subfigure} 

\usepackage{natbib}
\usepackage{verbatim}

\usepackage{algorithm}
\usepackage{algorithmic}

\usepackage{hyperref}

\newcommand{\Prob}{\mathbb{P}}

\newcommand{\unwsum}[1]{w_{Σ}({#1})}
\newcommand{\card}[1]{\vert #1 \vert}

\newcommand{\indep}{\protect\mathpalette{\protect\independenT}{\perp}}\def\independenT#1#2{\mathrel{\rlap{$#1#2$}\mkern2mu{#1#2}}} %alternative: \!\perp\!\!\!\perp
\newcommand{\dep}{\not\indep}

\newcommand{\est}{\mathfrak{I}}
\newcommand{\esth}{\widehat{\est}}
\newcommand{\estn}[1]{\est_{\mathrm{{#1}}}}
\newcommand{\estev}{\mathfrak{Z}}

\newcommand{\SaS}{{X}} %sample set
\newcommand{\smp}{{x}} %sample
\newcommand{\param}{α}
\newcommand{\ciRv}{γ}
\newcommand{\conRv}{\phi}
\newcommand{\dat}{d}

\newcommand{\lhoodfact}[2]{\prod_{\dat_i \dep \ciRv_#1} P(\dat_i | \conRv', \ciRv^{(#2)}_#1, \param_\dat) }
\usepackage{amsthm}

\newtheorem{dfn}{Definition}
\newtheorem{theo}{Theorem}
\begin{document}

%\twocolumn[
%\icmltitle{Consistency of Importance Sampling estimates based on dependent sample sets
%and an application to models with factorizing likelihoods}

%\icmlauthor{Ingmar Schuster}{schuster@informatik.uni-leipzig.de}
%\icmladdress{Natural Language Processing Group\\University of Leipzig}

%\icmlkeywords{importance sampling, is, population monte carlo, pmc, machine learning, ICML}

%\vskip 0.3in
%]
\title[Dependent Importance Sampling and Sample Inflation]{Consistency of Importance Sampling estimates based on dependent sample sets
and an application to models with factorizing likelihoods}
\author{Ingmar Schuster}

\address[Ingmar Schuster]{Natural Language Processing Group, University of Leipzig}
\email{schuster@informatik.uni-leipzig.de}

\begin{abstract}
In this paper, I proof that Importance Sampling estimates based on dependent sample sets are consistent under certain conditions. This can be used to reduce variance in Bayesian Models with factorizing likelihoods, using sample sets that are much larger than the number of likelihood evaluations, a technique dubbed Sample Inflation. I evaluate Sample Inflation on a toy Gaussian problem and two  Mixture Models. 
\end{abstract} 

\maketitle

\section{Introduction}
This paper broadens the scope of the Importance Sampling estimator by providing proofs that under rather mild conditions, estimates based on dependent sample sets are still consistent. This can be used for variance reduction in certain models classes, namely those that exhibit a factorizing structure in their likelihoods. The paper proceeds as follows. In Section~2, standard Importance Sampling techniques as well as an iterated Imporance Sampling scheme, Population Monte Carlo, are reviewed. Section~3 first exemplifies which models qualify as having a factorizing structure and introduces Sample Inflation for these models. Sample Inflation is a technique to artificially blow up the number of samples gained from few likelihood evaluations, thus attaining a much larger set of dependent samples. In Section~4, I proof that Importance Sampling estimates based on dependent samples are consistent, i.e. converge to the integral we are trying to estimate. Section~5 reviews related work from the Population Monte Carlo literature. Finally, Section~6 evaluates Sample Inflation on both a Gaussian toy problem as well as two Dirichlet Mixture Model estimations. In the conclusion, I give directions for future work.

\section{Importance Sampling}
The Importance Sampling estimator approximates the mean (alternatively: integral, expected value) $H$ of some function $h$ with respect to some probability density $f$: 
\begin{align*}
H &= ∫f(\smp)h(\smp) \mathrm{d}\smp\\
& = \mathbb{E}_f(h(\smp))
\end{align*}
 This is achieved by sampling from an auxiliary proposal density $q$. Say we have acquired a sample set $\SaS$ from $q$. The Importance Sampling estimator is given by
\begin{equation}
\est(\SaS) = \frac{1}{\vert \SaS \vert} \sum_{\smp ∈ \SaS} w(\smp)h(\smp) \label{eq:is_est}
\end{equation}
where $w(\smp) = f(\smp)/q(\smp)$ is the weight function \citep{Robert1999}. It can be used in case $f$ is not given only proportionally but exactly and is a probability density (i.e. is non-negative and integrates to $1$). A necessary condition for Importance Sampling to be unbiased is that $q(\smp) > 0$ whenever $f(\smp)h(\smp) ≠ 0$. Its variance is given by $\textrm{var}_q(\est(\SaS)) = σ^2_q/\card{\SaS}$ \citep[see][]{Owen2013}. To ensure finite variance, $q$ has to have heavier tails than $f$ \citep{Robert1999}.

However, most times we can only compute $f$ proportionally, as the normalizing constant (also called evidence or marginal likelihood) is unknown. In particular, this is often the case in Bayesian Inference, where the posterior over random variables is given proportionally by the product of prior and likelihood terms. Here, the self-normalized Importance Sampling estimator 
\begin{equation}
\label{eq:is_n_est}
\estn{n}(\SaS) = \frac{1}{\unwsum{\SaS}} \sum_{\smp ∈ \SaS} w_u(\smp)h(\smp)
\end{equation}
can be used \citep{Robert1999}, where $w_u(\smp) = f(\smp)/q(\smp)$ is the unnormalized weight function and $\unwsum{\SaS} = \sum_{\smp ∈\SaS}w_u((\smp))$. A variance estimate is given by $\sum_{\smp ∈ \SaS} (w_u(\smp)/\unwsum{\SaS})^2(h(\smp)-\estn{n}(\SaS))^2$ \citep{Owen2013}. Both standard and self-normalized Importance Sampling are consistent as a direct consequence of the strong law of large numbers  \citep[see][]{Geweke1989}.

\subsection{Population Monte Carlo}
\label{sec:pmc}
I will use the Population Monte Carlo \citep[PMC;][]{Cappe2004} paradigm in one of the experiments in the evaluation section. As PMC is not well known in the Machine Learning community, I will introduce it here it in a very concise way. However, the reader might as well skip this section at first and come back to it before reading section~\ref{sec:relwork}. See \citet{Cappe2004} for a thorough introduction  to PMC and \citet{Douc2007,Marin2012,Iacobucci2010} for newer developments. 

The PMC method is based on the observation that proposal distributions for Importance Sampling can depend on previous samples without compromising the validity or (asymptotic) unbiasedness of the estimator \citep{Cappe2004}. PMC works by first generating a population of importance samples (hence the name) from a set of proposal distributions. In each new generation of samples, proposal distributions can be built on previous sample generations. To equalize samples, an Importance Resampling step is introduced whereby  each sample in the population is resampled with replacement with a probability proportional to its weight \cite{Rubin1987}. A detailed description is given in Algorithm~\ref{alg:pmc}. The essential feature of the Algorithm in it is step (a): for each sample and each generation, an individual proposal distribution can be used, the only restriction being that it might not depend on samples from the same generation. In its most naive version (which I will be using), PMC enables choosing choosing the proposal distributions for a new generation such that they are centered on samples from previous generations. Generally speaking, the aim when choosing proposal distributions is to minimize the variance of importance weights - thus avoiding infinite variance of the estimate.

\begin{algorithm}[tb]
   \caption{Population Monte Carlo Algorithm}
   \label{alg:pmc}
\begin{algorithmic}
   \STATE {\bfseries Input:} initial proposal densities, unnormalized density $f$, population size $p$, sample size $m$
    \STATE {\bfseries Output:} list of $m$ samples

   \STATE Initialize $S = List()$
   \FOR{$t=1$ {\bfseries to} $T$}
   	\STATE Initialize $P = List()$
	\STATE Initialize $W = List()$
   	\FOR{$i=1$ {\bfseries to} $p$}
		\STATE (a) select proposal distribution $q_{i,t}$
		\STATE (b) generate $\smp \sim q_{i,t}$ and append it to $S$
		\STATE ~~~append weight $f(\smp)/q_{i,t}(\smp)$ to $W$
   	\ENDFOR
	\STATE normalize $W$ to sum to $1$
	\STATE resample $p$ values from $P$ with replacement with\\
		      probability given by the corresponding value in $W$\\
		     and append samples to $S$
   \ENDFOR
   \STATE return $S$
\end{algorithmic}
\end{algorithm}

\section{Models with factorizing likelihood terms}
\label{sec:factor_posterior}
Assume our generative model has the following structure.
\begin{align*}
\conRv &\sim P(\conRv | \param_\conRv) &{ }\\
\ciRv_j &\sim P(\ciRv_j|\param_\ciRv)~&∀ j ∈ [1, …,K]\\
\dat_i &\sim P(\dat_i | \conRv, \ciRv, \param_\dat)& ∀ i ∈ [1, …,N]
\end{align*}

where $\dat_i$ is the $i$th data point, there are $N$ data points,  each $P$ represents some parameterized family of distributions and $\param_\conRv, \param_\ciRv, \param_\dat$ are fixed model parameters. Then the posterior over the latent variables $\conRv, \ciRv$ is given by
\begin{equation*}
p(\conRv, \ciRv | d) \propto P(\conRv | \param_\conRv) \prod_{j=1}^K P(\ciRv_j|\param_\ciRv)  \prod_{i=1}^N P(\dat_i | \conRv, \ciRv, \param_\dat)
\end{equation*}
where $N$ is the number of data points. Now assume further that the likelihood term for each data point $\dat_i$ depends exactly on one $\ciRv_j$ ( $\dat_i \dep \ciRv_j$) and is independent of the other variables in $\ciRv$ ( $\dat_i \indep \ciRv_m$ for $m ≠ j$). %\textcolor{green}{This is not a general result of the V-structure I introduced - the general expression $P(\dat_i | \conRv, \ciRv, \param_\dat)$ allows for $\dat_i$ to depend on all $\ciRv$, so we have to state independence explicitly}
This induces a partition on the data points and allows for further factorization of the likelihood term
\begin{equation*}
  \prod_{i=1}^N P(\dat_i | \conRv, \ciRv, \param_\dat) = \prod_{j=1}^K \prod_{\dat_i \dep \ciRv_j} P(\dat_i | \conRv, \ciRv_j, \param_\dat)
\end{equation*}
This model structure renders the individual $\ciRv_i$ conditionally independent of each other, 
\begin{equation}
\ciRv_i \indep  \ciRv_j | \param, \conRv, \dat~\textrm{for}~i ≠j
\end{equation}
 Two model classes satisfying these assumptions are probabilistic matrix factorization (discussed in \ref{sec:mf}) and Dirichlet Mixture Models (discussed in \ref{sec:dmm}). First however, I will exemplify an Importance Sampling method, called Sample Inflation, that is applicable whenever the assumptions above hold.

\subsection{Sample Inflation for Importance Sampling}
A straight forward self-normalized Importance Sampler for models with factorizing likelihoods is given in Algorithm~\ref{alg:isIID}. If the density $f$ is actually given in normalized form, the self-normalization at the end ($W = W / (\sum_{w ∈ W} w)$) can be skipped.
\begin{algorithm}[tb]
   \caption{Importance Sampling for factorizing models }
   \label{alg:isIID}
\begin{algorithmic}
   \STATE {\bfseries Input:} proposal densities $q_\conRv,q_{\ciRv_1}, …,q_{\ciRv_K}$, unnormalized density $f$, sample size $m$
    \STATE {\bfseries Output:} tuple $(S,W)$ of $m$ samples and weights
   \STATE Initialize samples list $S = List()$
   \STATE Initialize weights list $W = List()$
   \WHILE{len(S) $< m$}
       \STATE sample $\conRv'$ according to $q_\conRv$
       \FOR{$j=1$ {\bfseries to} $K$}
           \STATE sample $\ciRv'_j$ according to $q_{\ciRv_j}$
       \ENDFOR
       \STATE append $(\conRv', \ciRv'_1, …,\ciRv'_K)$ to $S$
       \STATE append ${f(\conRv', \ciRv'_1, …,\ciRv'_K)}/{  ( q_\conRv(\conRv') \prod_{j=1}^K q_{\ciRv_j}(\ciRv'_j) )}$ to $W$
   \ENDWHILE
   \STATE $W = W / (\sum_{w ∈ W} w)$
   \COMMENT{for self-normalized IS}
\end{algorithmic}
\end{algorithm}

Now consider the following modification of Algorithm~\ref{alg:isIID}: instead of only generating one sample $\ciRv'_j$ from $q_{\ciRv_j}$, generate two samples $\ciRv^{(1)}_j, \ciRv^{(2)}_j$ and append both $(\conRv', \ciRv^{(1)}_1, …,\ciRv^{(1)}_K)$ and $(\conRv', \ciRv^{(2)}_1, …,\ciRv^{(2)}_K)$ to the sample list $S$ (and the accompanying weights to the weight list $W$). Contrary to first intuition, a set of samples generated this way does not jeopardize consistency, for the corresponding proof see section~\ref{sec:didConsistency}. The likelihood term for the second sample costs as much to compute as the likelihood term for the first sample. As the likelihood term is usually the most expensive part of posterior computation, we get two dependent samples (because the same $\conRv'$ appears in both of them) for the computational price of two independent samples. However, we can take advantage of the likelihood structure to get an overall of $2^K$ dependent samples. If we sample $M$ initial samples instead, we can construct $M^K$ dependent samples for the price of $M$ likelihood evaluations. This grows very quickly, in fact the growth is polynomial in $M$ and exponential in $K$.

For ease of illustration, consider $M=2, K=2$. The likelihood term for the first sample  $(\conRv', \ciRv^{(1)}_1, \ciRv^{(1)}_2)$ is

\begin{equation*}
\lhoodfact{1}{1}   \lhoodfact{2}{1}
\end{equation*}
and for $(\conRv', \ciRv^{(2)}_1, \ciRv^{(2)}_2)$  we have the likelihood
\begin{equation*}
\lhoodfact{1}{2}   \lhoodfact{2}{2}
\end{equation*}
Reusing the factors computed for the first two samples, we can calculate the likelihoods  of two more dependent samples, $(\conRv', \ciRv^{(1)}_1,\ciRv^{(2)}_2)$ and  $(\conRv', \ciRv^{(2)}_1,\ciRv^{(1)}_2)$,  almost for free! \\
This gives rise to the Sample Inflation method, given in Algorithm~\ref{alg:isSI}. In the algorithm, I use  $c$ as a shorthand ranging over joint samples for the random variables $\ciRv_1,…,\ciRv_K$ and $q_\ciRv(c)$ as a shorthand for $q_{\ciRv_1}(c_1), …,q_{\ciRv_K}(c_K)$. A way to think about Sample Inflation is that we can use the structure of the problem to get a better approximation of the marginal $f(\conRv)$ by averaging over an inflated sample set for $\ciRv$. %FI\SaSME: when accepted, thank Christian Robert for this interpretation
\begin{algorithm}[tb]
   \caption{Importance Sampling (Sample Inflation)}
   \label{alg:isSI}
\begin{algorithmic}
   \STATE {\bfseries Input:} proposal densities $q_\conRv,q_{\ciRv_1}, …,q_{\ciRv_K}$, unnormalized density $f$, number of independent proposals for $\conRv$ $m$, number of likelihood evaluations per independent proposal of $\conRv$ $M$
    \STATE {\bfseries Output:} tuple $(S,W)$ of $m \cdot M^K$ samples and weights

   \STATE Initialize $S = List()$
   \STATE Initialize $W = List()$
   \WHILE{len(S) $< m$}
       \STATE sample $\conRv'$ according to $q_\conRv$
       \FOR{$j=1$ {\bfseries to} $K$}
         \FOR{$i=1$ {\bfseries to} $M$}
             \STATE sample $\ciRv^{(i)}_{j}$ according to $q_{\ciRv_{j}}$
       	\ENDFOR
       \ENDFOR
       \STATE compute set $C$ of all $M^K$ possible joint samples 
       		from set of tuples $\{(\ciRv^{(m)}_1,…,\ciRv^{(m)}_K) :m ∈ \{1,…,M\}\}$
       \FOR{$c ∈ C$}
         \STATE append $(\conRv', c)$ to $S$
         
         \STATE append $f(\conRv', c) / \left ( q_\conRv(\conRv') q_{\ciRv}(c) \right)$ to W
         \COMMENT{reuse previous likelihood factor computations for $f$}
        \ENDFOR
   \ENDWHILE
   \STATE $W = W / (\sum_{w ∈ W} w)$
   \STATE return $(S, W)$
\end{algorithmic}
\end{algorithm}

\subsection{Matrix Factorization}
\label{sec:mf}
For illustration purposes I will discuss Factor Analysis. Other examples of Bayesian matrix factorization models include \emph{Gamma Process Nonnegative Matrix Factorization} \cite{Hoffman2010}, \emph{ Probabilistic Matrix Factorization} \cite{Salakhutdinov2007} and Poisson Factorization \cite{Gopalan2013}. The Factor Analysis model with $k$ latent factors has the structure
\begin{equation*}
\dat_i = \conRv \ciRv_i + ε_i
\end{equation*}
Here $\dat_i ∈ \mathbb{R}^p$, $ε_i \sim N(0, Σ)$ is a residual for some covariance matrix $Σ$, $\conRv ∈ \mathbb{R}^{p\times k}$ is a factor loading matrix and $\ciRv_i \sim N(0, I_k)$ is a vector of latent factors (one for each data point, thus $K$ equals the number of data points). I will not discuss the choice of priors on $\conRv$ and $Σ$; for a profound discussion of Factor Analysis, see \citet{Dunson2006}. The key observation is that the likelihood of $\dat_i$ does not depend on $\ciRv_j$ for $j ≠ i$ and thus teach $\ciRv_j$ is conditionally independent of all the other variables in $\ciRv$: $\ciRv_i \indep  \ciRv_j | \param, \conRv, \dat$ for $i ≠j$. Thus, the assumptions from section \ref{sec:factor_posterior} are satisfied. Sample inflation in the case of factor analysis works by first sampling proposals $\conRv'$ and possibly $Σ'$, then sampling $M$ proposals for each $\ciRv_i$. The likelihood of a single sample $\ciRv^{(1)}_i$ then is evaluated as $N(\dat_i | \conRv'\ciRv^{(1)}_i, Σ')$.
Lets say we have two data points $\dat_1, \dat_2$ and two samples for each of the $\ciRv_i$. The likelihood of the two joint samples $(\conRv',Σ',\ciRv^{(1)}_1, \ciRv^{(1)}_2), (\conRv',Σ',\ciRv^{(2)}_1, \ciRv^{(2)}_2)$ is
\begin{align*}
N(\dat_1 | \conRv'\ciRv^{(1)}_1, Σ')& N(\dat_2 | \conRv'\ciRv^{(1)}_2, Σ')\\
\textrm{and}\\
N(\dat_1 | \conRv'\ciRv^{(2)}_1, Σ')& N(\dat_2 | \conRv'\ciRv^{(2)}_2, Σ').
\end{align*}
From the factors computed for these two samples, we get the likelihood for $(\conRv',Σ',\ciRv^{(1)}_1, \ciRv^{(2)}_2)$ and $(\conRv',Σ',\ciRv^{(2)}_1, \ciRv^{(1)}_2)$ using almost no additional computation time. In general we get $M^K$ samples using $M$ likelihood evaluations.

\subsection{Dirichlet Mixture Models}
\label{sec:dmm}
In Dirichlet Mixture Models each data point is assumed to be generated by a mixture of $K$ base distributions, where parameters of the base distributions are given by $\ciRv_1,…, \ciRv_K$. A Dirichlet prior is placed on the mixture proportions $\conRv^{(1)}$. For each data point $\dat_i$ a categorical variable $\conRv_i^{(2)}$ is drawn, indicating which base distribution it is generated from.\footnote{The notation differs from the usual notation in DP Mixture Models. However, I valued consistency with section~\ref{sec:factor_posterior} higher than consistency with the rest of the literature.} The full generative model is
\begin{align*}
\ciRv_j &\sim G_0(\param_\ciRv) & ∀ j ∈ \{1,…,K\}\\
\conRv^{(1)} &\sim \textrm{Dir}(\param_\conRv) \\
\conRv^{(2)}_i &\sim \textrm{Cat}( \conRv^{(1)} ) \\
\dat_i &\sim P(\dat_i | \ciRv_{\conRv^{(2)}_i}) 
\end{align*}

where $G_0$ is a prior on the parameters of the $K$ base distributions, $\conRv^{(1)} ∈ \mathbb{R}_+^K$, $\conRv^{(2)}_i ∈ \{1,…,K\}$ and $P(\cdot | \ciRv_{\conRv^{(2)}_i})$  is the base distribution with index $\conRv^{(2)}_i$ (each base distribution could also have some global parameter $\param_\dat$, which I drop for notational clarity). Again, observe that $d_i$  does not depend on $\ciRv_j$ for $j ≠ \conRv^{(2)}_i$ and the assumptions from section \ref{sec:factor_posterior} hold because $\ciRv_i \indep  \ciRv_j | \param_\ciRv, \param_\conRv, \conRv^{(1)}, \conRv^{(2)}, \dat$ for $i ≠j$. To apply Sample Inflation to Dirichlet Mixture Models, one would first sample a proposal $\conRv'^{(1)}$ and $\conRv'^{(2)}_i$ for each $i$, then $M$ proposals for each $\ciRv_j$, and recombine these to get $M^K$ dependent samples.
  
\section{Related Work}
\label{sec:relwork}
To the best of my knowledge, a recombination of Importance Samples as suggested in this paper has not been proposed before.\\ 
Generally speaking, variance reduction is an important topic in Importance Sampling and its descendant Population Monte Carlo. I will concentrate on the PMC case here. In the original paper by \citet{Cappe2004}, the approach used for variance reduction is to keep several markov transition kernels which generate new samples centered on previous ones with a different variance for each kernel. Those kernels which exhibit smaller weight variance are then used more often. Mixture-PMC  \citep[M-PMC; ][]{Cappe2008} goes one step further in that it fits a Gaussian or Multivariate $t$ mixture model to the samples from previous generations, generating new samples from this approximation of the posterior. D-Kernel PMC by \citet{Douc2007} fits a D-Kernel Mixture and can be shown to converge to the optimum D-Kernel Mixture.

\section{{I}mportance {S}ampling estimators based on dependent samples}
\label{sec:didConsistency}
In the literature, the sample set used for the Importance sampling estimator is often assumed to consist only of independently identically distributed (iid) samples. However, one potentially interesting (and as we will see practically very relevant) case is when samples are guaranteed to come from the proposal density $q$ but are not required to be independent. I will first introduce some assumptions and notation for this section.

\begin{dfn} Let $\SaS_1, …, \SaS_k$ with fixed $k$ be (multi-)sets of samples from some density (for claims about the Importance Sampling estimator, from the proposal density $q$). The samples in each $\SaS_i$ are assumed to be iid but the samples in the (multi-)set $\SaS_\cup =  \bigcup_i \SaS_i$ are not necessarily  independent. Furthermore, let $\SaS^{(m)}_\cup = \bigcup_{i=1}^k \SaS^{(m)}_i$ be a sequence of sample sets with fixed $k$, $m = \left \vert \SaS^{(m)}_\cup \right \vert$ and $|\SaS^{(m)}_i|\stackrel{m\to\infty}\longrightarrow \infty$. The samples in each $\SaS^{(m)}_i$ are assumed to be iid for any $m$ and $i$, but the samples in $\SaS^{(m)}_\cup$ might be dependent.
\end{dfn}

Now as a first step towards proving consistency of Importance Sampling estimates based on dependent sample sets, we note that the normed error of any convex combination of estimates based on iid sample sets cannot increase compared to the same convex combination of normed errors of individual estimates.

\begin{theo} \label{theo:is_cvx_estimator}
Let $\esth$ be any estimator of the true quantity $H$. Then the normed error of a convex combination of estimates $\sum_{i=1}^k λ_i\esth(\SaS_i)$ cannot exceed the convex combination of normed errors: 
\begin{equation*}
  \sum_{i=1}^k λ_i \| \esth(\SaS_i) -H \| ≥ \| \sum_{i=1}^k λ_i\esth(\SaS_i)-H \| ≥ 0
 \end{equation*}
 for any norm $\|\cdot\|$ and $\sum_{i=1}^k λ_i =1, ∀~i:λ_i ≥ 0$. In particular, this implies the squared error of the convex combination of estimators cannot exceed the convex combination of squared errors.
 \end{theo}
\begin{proof}
We have
\begin{align*}
&{ }&  \sum_{i=1}^k λ_i\| \esth(\SaS_i) -H \|\\
& ≥ &  \left \| \sum_{i=1}^k λ_i ( \esth(\SaS_i) -H ) \right \| \\
& = &  \left \| \left ( \sum_{i=1}^k λ_i  \esth(\SaS_i) \right ) - \left ( \sum_{i=1}^k λ_i H  \right ) \right \| \\
& = &  \| \sum_{i=1}^k λ_i\esth(\SaS_i)-H \| \\
& ≥ & 0
\end{align*}
where first inequality follows either from subadditivity and absolute homogeneity of norms or from Jensens inequality and the fact that norms are convex. The second inequality follows from the positivity property of norms.
\end{proof}

Now I will specialize Theorem~\ref{theo:is_cvx_estimator} to the case of the (normalized) Importance Sampling estimator. Recall that we are trying to estimate the integral  $H = ∫f(\smp)h(\smp) \mathrm{d}\smp$.

\begin{theo} \label{theo:is_se}
The normed error of the estimate $\est(\SaS_\cup)$ cannot exceed the cardinality weighted average of normed errors:
\begin{equation*}
 \sum_{i=1}^k \frac{\vert \SaS_i \vert}{\vert \SaS_\cup \vert} \| \est(\SaS_i) - H \| ≥ \| \est(\SaS_\cup) -H \| ≥ 0
 \end{equation*}
 where $\card\cdot$ signifies the cardinality of a set.\\
 Furthermore, the normed error of the estimate $\estn{n}(\SaS_\cup)$ cannot exceed the importance weighted average of normed errors: \begin{equation*}
 \sum_{i=1}^k \frac{\unwsum{\SaS_i}}{\unwsum{\SaS_\cup}} \| \estn{n}(\SaS_i) -H \| ≥ \| \estn{n}(\SaS_\cup) -H \| ≥ 0.
 \end{equation*}
\end{theo}
\begin{proof}
\newcommand{\saSetSize}[1]{\card{#1}}
\newcommand{\estTmp}{\est}
\newcommand{\estimatorDefinition}{\eqref{eq:is_est} }
\newcommand{\wTmp}{w}
For the case of the unnormalized estimator 
%%begin copied from area %%
$\estTmp$, if we choose $λ_i = \saSetSize{\SaS_i}/\saSetSize{\SaS_\cup}$ and show  $ \sum_{i=1}^k \frac{\saSetSize{\SaS_i}}{\saSetSize{\SaS_\cup}} \estTmp(\SaS_i) =  \estTmp(\SaS_\cup)$, the claim follows from Theorem~\ref{theo:is_cvx_estimator}. Using the definition of the estimator \estimatorDefinition we have
\begin{align*}
 \sum_{i=1}^k \frac{\saSetSize{\SaS_i}}{\saSetSize{\SaS_\cup}}   \estTmp(\SaS_i) &\stackrel{\estimatorDefinition}{=} & 
 \sum_{i=1}^k \frac{1}{\saSetSize{\SaS_\cup}}   \sum_{\smp ∈ \SaS_i} \wTmp(\smp)h(\smp)\\
&= &  \frac{1}{\saSetSize{\SaS_\cup}}   \sum_{\smp ∈ \SaS_\cup} \wTmp(\smp)h(\smp)\\
& \stackrel{\estimatorDefinition}{=}  &  \estTmp(\SaS_\cup)
\end{align*}
%%end copied from area%%
and thus the first claim holds.
\renewcommand{\saSetSize}[1]{\unwsum{#1}}
\renewcommand{\estTmp}{\estn{n}}
\renewcommand{\estimatorDefinition}{\eqref{eq:is_n_est} }
\renewcommand{\wTmp}{w_u}
For the case of the self-normalized estimator
%%begin copied to area %%
$\estTmp$, if we choose $λ_i = \saSetSize{\SaS_i}/\saSetSize{\SaS_\cup}$ and show  $ \sum_{i=1}^k \frac{\saSetSize{\SaS_i}}{\saSetSize{\SaS_\cup}} \estTmp(\SaS_i) =  \estTmp(\SaS_\cup)$, the claim follows from Theorem~\ref{theo:is_cvx_estimator}. Using the definition of the estimator \estimatorDefinition we have
\begin{align*}
 \sum_{i=1}^k \frac{\saSetSize{\SaS_i}}{\saSetSize{\SaS_\cup}}   \estTmp(\SaS_i) &\stackrel{\estimatorDefinition}{=} & 
 \sum_{i=1}^k \frac{1}{\saSetSize{\SaS_\cup}}   \sum_{\smp ∈ \SaS_i} \wTmp(\smp)h(\smp)\\
&= &  \frac{1}{\saSetSize{\SaS_\cup}}   \sum_{\smp ∈ \SaS_\cup} \wTmp(\smp)h(\smp)\\
& \stackrel{\estimatorDefinition}{=}  &  \estTmp(\SaS_\cup)
\end{align*}
%%end copied to area%%
and thus the second claim holds.
\end{proof}

To get an intuition for the meaning of Theorem \ref{theo:is_se} for the case of the unnormalized estimator, recall that by using Algorithm~\ref{alg:isSI}, we can get $M^K$ iid sample sets. Each of these is of size $m$, so the convex combination amounts to a simple average. Thus, we can only do better on average by using the samples from all sets as compared to the samples from only one set. This seems particularly fortunate after realizing that there is no reason to prefer one of the sample sets over one of the others (all of them are sampled iid from $q$).

Now if the normed error cannot increase when using dependent sample sets for estimation, we might expect that the estimate converges in probability to the true integral $H$. In other words, we might expect that the sequence of estimates is consistent. This is indeed the case as stated by the following theorem.

\begin{theo} \label{theo:is_consist}
Let $\esth$ be the (self-normalized) Importance Sampling estimator. The sequence of estimates $\esth(\SaS^{(m)}_\cup)$ for $m→\infty$ is consistent, i.e.
\begin{equation*}
\lim_{m → \infty} \Prob(\|\esth(\SaS^{(m)}_\cup) -H \| ≥ ε) = 0
\end{equation*}
for all $ε > 0$.
\end{theo}
\begin{proof}
By standard results for all $i$ and any $ ε > 0$: $$\lim_{\vert \SaS^{(m)}_i \vert → \infty} \Prob(\|\esth(\SaS^{(m)}_i) -H \| ≥ ε) = 0 $$
Now assume that This implies that for any convex combination
\begin{equation*}
 \lim_{m → \infty} \Prob\left (\sum^k_{i=1} λ^{(m)}_i \|\esth(\SaS^{(m)}_i) -H \| ≥  ε \right ) = 0
\end{equation*}
using $\sum_i λ^{(m)}_i ε = ε$. Now if $\esth = \est$ choose  $λ^{(m)}_i = {| \SaS^{(m)}_i |}/{| \SaS^{(m)}_\cup |}$ 
, if $\esth = \estn{n}$ choose  $λ^{(m)}_i = {\unwsum{\SaS^{(m)}_i}}/{\unwsum{\SaS^{(m)}_\cup}}$ and apply Theorem \ref{theo:is_se}  to get
\begin{equation*}
\lim_{m → \infty} \Prob(\|\esth(\SaS^{(m)}_\cup) -H \| ≥ ε) = 0
\end{equation*}
%\textcolor{blue}{(You did not treat the case in which $\vert \SaS^{(m)}_i \vert → \infty$ does not hold for some $i$. You might consider to exclude this case from the statement of your theorem in the first place (by replacing $\SaS^{(m)}_i \subseteq \SaS^{(m+1)}_i$ by $|\SaS^{(m)}_i|\stackrel{m\to\infty}\longrightarrow \infty$. Otherwise you should treat this case rigorously, which is not difficult but cumbersome)}
 \end{proof}

%\textcolor{blue}{Maybe the following summary of the proof appears helpful in some way...:
%\\
%Define the events
%\[
%A_i^{(m)} = \|\esth(\SaS^{(m)}_i) -H \|
%\ \ \text{and}\ \ 
%B^{(m)} = (\|\esth(\SaS^{(m)}_\cup) -H \|
%\]
%Then we have
%\[
%B^{(m)}≥ ε \stackrel{\text{Theorem 1 or 2}}{\Longrightarrow}
%\sum_{i=1}^k \lambda_i^{(m)} A_i^{(m)} ≥ ε \Rightarrow
%\exists i:\ A_i^{(m)}≥ ε.
%\]
%As a consequence
%\begin{align*}
%\Prob\left(B^{(m)}≥ ε\right)
%&\le
%\Prob\left(\sum_{i=1}^k \lambda_i^{(m)} A_i^{(m)} ≥ ε\right)
%\\ &\le
%\Prob\left(\exists i:\ A_i^{(m)}≥ ε\right)
%\\ &\le
%\sum_{i=1}^k\Prob\left(A_i^{(m)}≥ ε\right)
%\stackrel{m\to\infty}{\longrightarrow} 0
%\end{align*}
%}

An important detail of Theorem \ref{theo:is_consist} is that the number of sets $k$ is fixed as $m → \infty$.

One of the major reasons for choosing Importance Sampling over other simulation techniques is that it enables approximating model evidence (also called marginal likelihood or the normalizing constant of $f$). This is based on the identity
\begin{equation*}
F =∫f(\smp)\textrm{d}\smp 
= ∫\frac{f(\smp)}{q(\smp)}q(\smp)\textrm{d}\smp
= \mathbb{E}_q \left( \frac{f(\smp)}{q(\smp)} \right )
\end{equation*}

which yields the unbiased and consistent estimator
\begin{align}
\estev(\SaS) &= \frac{1}{|\SaS|} \sum_{\smp ∈ \SaS} f(\smp)/q(\smp)
\end{align}
Evidence estimates based on sample sets that contain dependent samples will stay consistent as follows from Theorem \ref{theo:is_consist} by setting $h(\smp) = 1$.

%\textcolor{blue}{(For a mathematical paper, you should also state which spaces your functions live in. What space are they defined on (e.g. $f,h\colon \mathbb R^d\to\mathbb R$)? Are they continuous ($C(\mathbb R^d)$) / integrable ($L^1(\mathbb R^d)$) / square-integrable ($L^2(\mathbb R^d)$ / differentiable...? Also you should support the results you use by citing other papers/books. Also some citing in the introduction would be nice together with a description of the "state of the art". As I said, this only holds for mathematical papers.)}

\section{Evaluation}
In this section, I will evaluate Sample Inflation for two cases. As a very simple measure, we will look into the performance of Sample Inflation when computing the (known) expectation of a two dimensional Gaussian distribution with diagonal covariance matrix. As a more involved case we will consider the estimation of two Dirichlet Mixture Models.

\subsection{Expectation of a multivariate Gaussian}
\begin{figure*}[ht]
\vskip 0.2in
\begin{center}
\centerline{\includegraphics[width=\textwidth ]{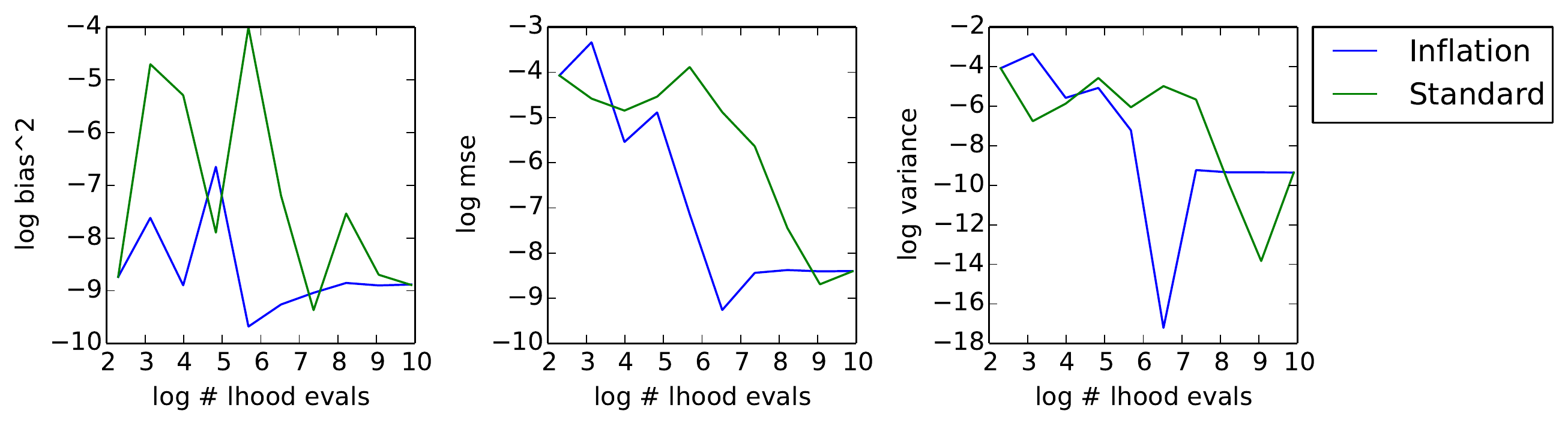}}
\centerline{\includegraphics[width=\textwidth]{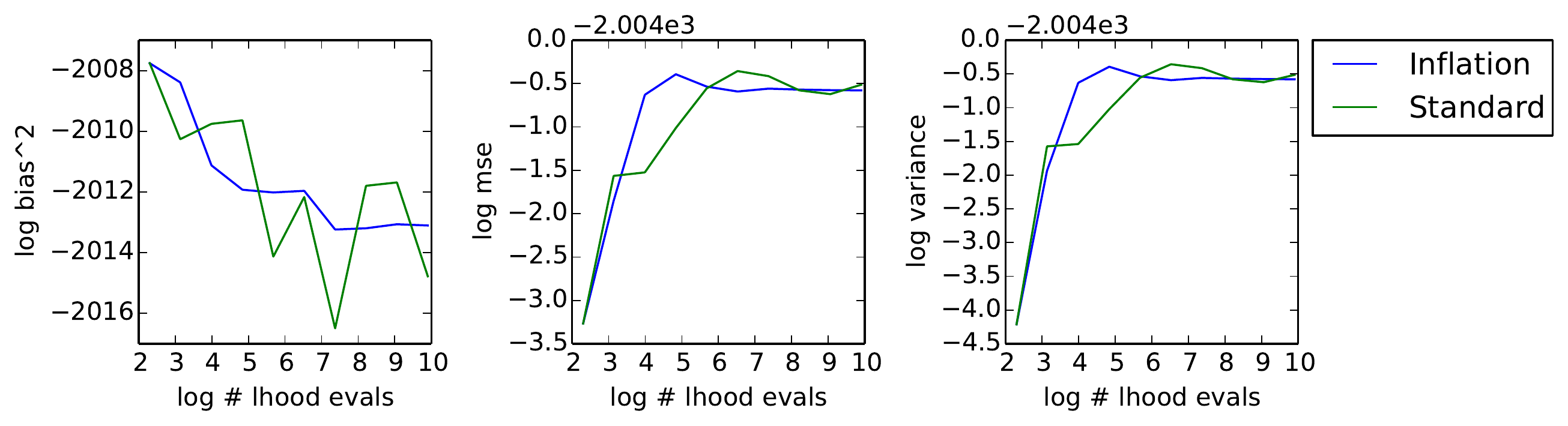}}
\caption{Performance of Sample Inflation as compared to standard self-normalized Importance Sampling from the same proposal distribution. The top plots give squared bias, mean squared error and variance for estimation of the expectation of the target distribution.The bottom plots plot the same measures for estimation of the evidence. The proposal distribution was centered on the true expectation of the target distribution.}
\label{fig:simple_gauss}
\end{center}
\vskip -0.2in
\end{figure*}

For the two experiments in this subsection, $20,000$ samples where drawn from the respective two dimensional proposal distribution. These were unchanged for standard Importance Sampling estimation. For Sample Inflation, the sample set was partitioned into sets of $100$ samples, which where inflated and concatenated into $2,000,000$ dependent samples.\\
As a first evaluation case I chose  a multivariate normal, $f = N(0, 2I)$, as the target distribution. The log evidence (log normalizing constant) was artificially set to $-1000$. The proposal distribution was a multivariate $t$-Distribution with the same mean and covariance matrix and $20$ degrees of freedom, $q = T(0,2I, 20)$. Squared bias, mean squared error (MSE) and variance of the estimates for the expectation of the target distribution as well as the evidence are given in log-log-plots in Figure~\ref{fig:simple_gauss}. For estimation of the targets expectation, the MSE, which subsumes variance and squared bias, clearly shows that using Sample Inflation is preferable to standard Importance Sampling.  The picture is less clear cut for evidence approximation, but  Sample Inflation does not seem hurt performance strongly.

\begin{figure*}[ht]
\vskip 0.2in
\begin{center}
\centerline{\includegraphics[width=\textwidth]{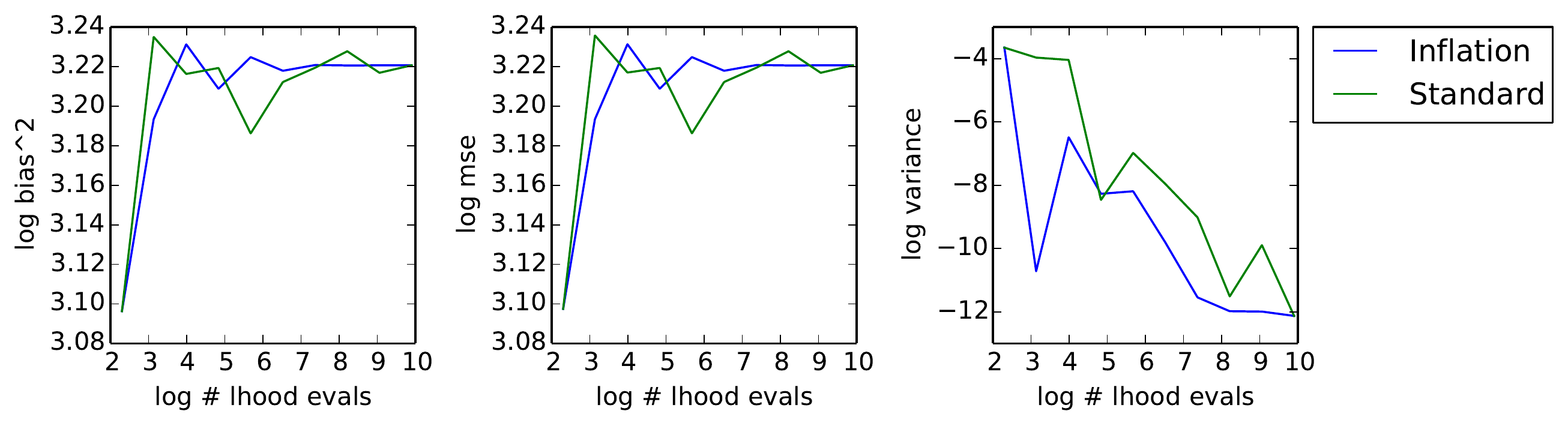}}
\centerline{\includegraphics[width=\textwidth]{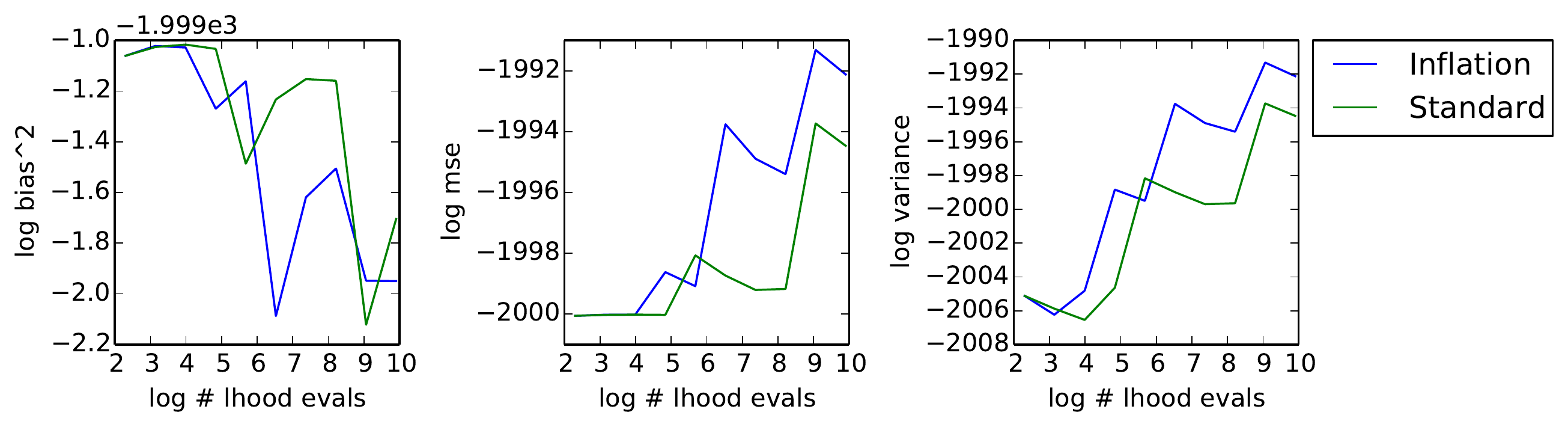}}
\caption{Performance of Sample Inflation as compared to standard self-normalized Importance Sampling from the same proposal distribution. The top plots give squared bias, mean squared error and variance for estimation of the expectation of the target distribution.The bottom plots plot the same measures for estimation of the evidence. The proposal distribution was \emph{not} centered on the true expectation of the target distribution.}
\label{fig:off_gauss}
\end{center}
\vskip -0.2in
\end{figure*}

The second experiment used the same target, $f = N(0, 2I)$ with a log evidence of $-1000$. This time however, the proposal distribution was not centered on the target, but on $(5,5)^T$, $q = T((5,5)^T,2I, 20)$. The mean squared error evaluation does not favor Sample Inflation for estimation of the targets expectation this time, though Sample Inflation gives more stable estimates (Figure~\ref{fig:off_gauss}). The major contribution to MSE here comes from the bias, which is caused by the fact that our proposal distribution is not centered on the target. For evidence approximation, Sample Inflation hurts performance slightly, but bear in mind that the differences to standard Importance Sampling are small when transformed back from log space.

\subsection{Estimation of DMMs}
\begin{figure*}[ht]
\vskip 0.2in
\begin{center}
\centerline{\includegraphics[width=\textwidth]{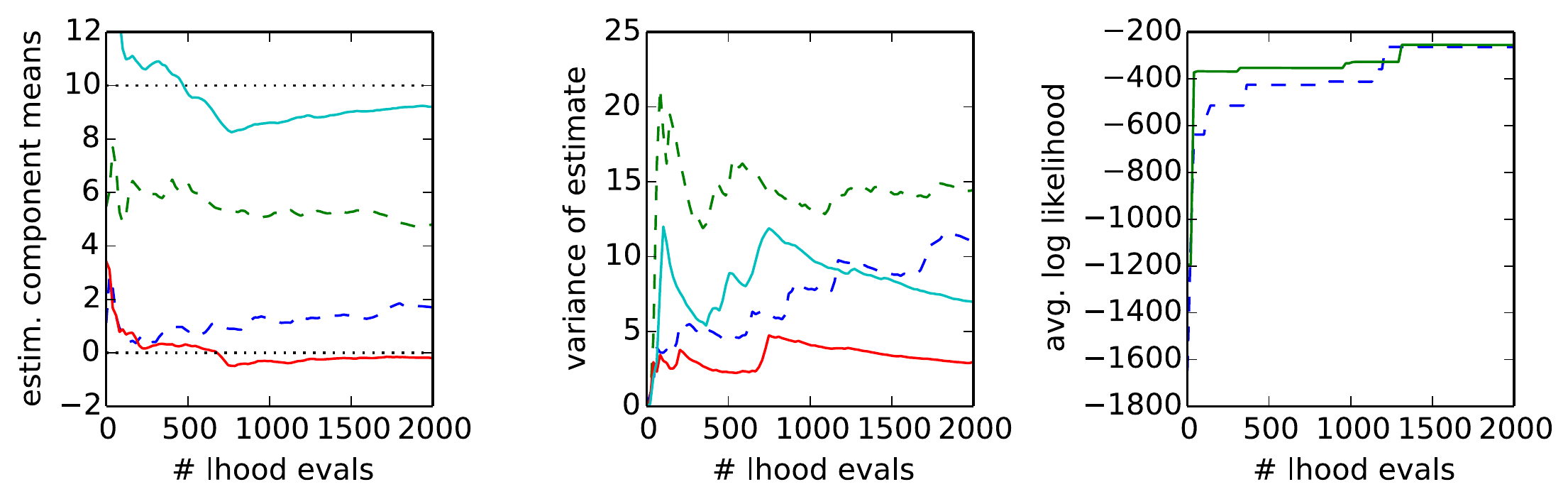}}
\caption{Gaussian Mixture Model estimation. Solid lines mark Sample Inflation, dashed lines standard Importance Sampling. The true means of the synthetic data are dotted. Using Sample Inflation, high likelihood regions are reached more quickly. This is reflected in the estimated means, which are closer to the true means of the data for Sample Inflation.}
\label{fig:gmm}
\end{center}
\vskip -0.2in
\end{figure*}

In this evaluation, I use a Population Monte Carlo approach to estimate two Dirichlet Mixture Models (DMMs) for synthetic data sets comprised of $100$ data points. For both experiments, $2000$ samples where drawn for standard Importance Sampling. For Sample Inflation, after sampling $\conRv^{(1)}$ and $\conRv^{(2)}$, two dependent samples where drawn for the parameters of the two component distributions (thus $M=2, K=2$). I used less overall samples for Sample Inflation than for standard Importance Sampling, so as to keep the number of likelihood evaluations exactly equal.\\
In the first case, the synthetic data was generated from a mixture of two one dimensional Gaussians with different means and unit variance. The DMM used two Gaussian components with fixed unit variance. Thus, only the means of the components had to be estimated. I put an $N(0,1)$ prior on the component means. I used Gaussian Markov kernels to generate proposals based on samples from earlier generations of the PMC algorithm. Sample Inflation attained regions of high likelihood more quickly and exhibited lower variance than standard Importance Sampling (Figure~\ref{fig:gmm}).

\begin{figure*}[ht]
\vskip 0.2in
\begin{center}
\centerline{\includegraphics[width=\textwidth]{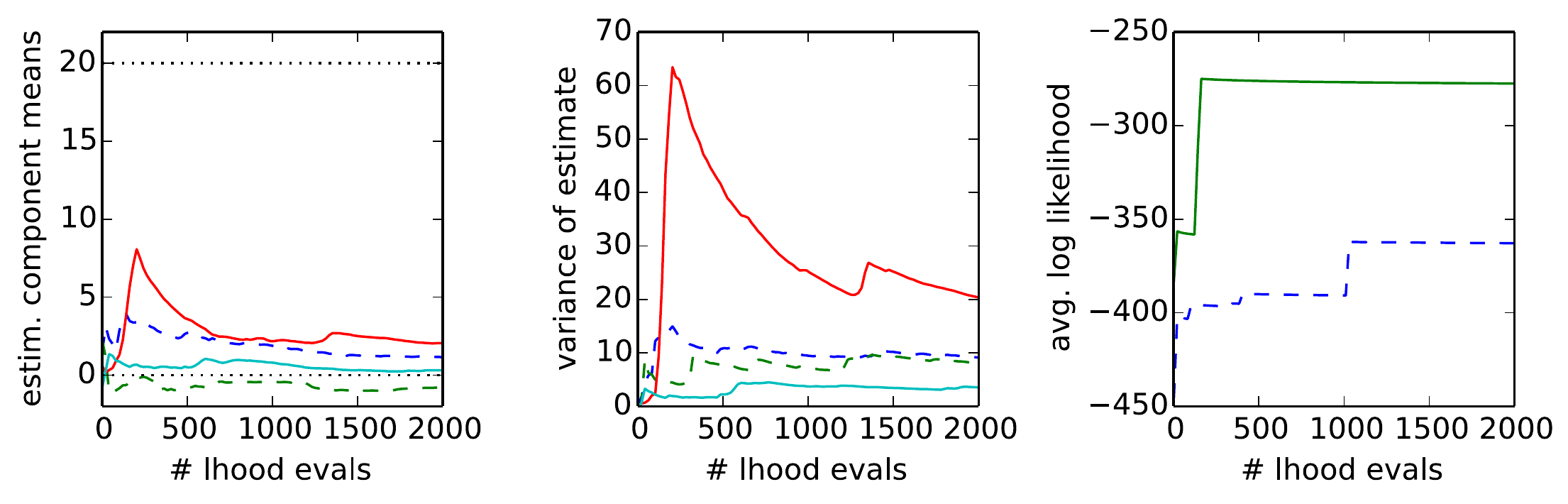}}
\caption{T Mixture Model estimation. Solid lines mark Sample Inflation, dashed lines standard Importance Sampling. The true means of the synthetic data are dotted. Using Sample Inflation, high likelihood regions are reached much more quickly.}
\label{fig:Tmm}
\end{center}
\vskip -0.2in
\end{figure*}

In the second case, the synthetic data was generated from a mixture of two one dimensional T-Distributions with different means, unit variance, and $30$ degrees of freedom. The DMM used two T components. Based on a sample $S$ from a previous generation, I used T-distributed Markov kernels to generate proposals for the mean centered on value of the mean parameter in $S$. Equivalently, I used Inverse-Wishart Distributions centered on the covariance matrix in $S$ and Gamma distributions centered on the degrees of freedom in $S$. A  Student-$t$ $T(0,1, 1)$ prior was placed on the component means, an Inverse Wishart
 $IW(σ^2=5, \textrm{df}=1)$ prior on the covariance and a $Gamma(1,1)$ prior (shape and scale parametrization) on the degrees of freedom. Here, Sample Inflation is much better in achieving high likelihoods more quickly, though the estimates exhibit higher variance (Figure~\ref{fig:Tmm}). The estimates of the means are not close to the true means of the synthetic data, which probably 
 stems from the fact that the Mixture Model is very flexibly as we also estimate the covariance matrix and degrees of freedom. Also, the $IW(σ^2=5, \textrm{df}=1)$ prior on the covariance and the $Gamma(1,1)$ prior on degrees of freedom are very broad and compensate easily for the rather narrow $T(0,1, 1)$ prior on the component means.

\section{Conclusion}
The contributions of this paper where twofold. First, I proved that Importance Sampling estimates based on dependent sample sets are consistent under mild conditions. To the best of my knowledge, this has not been proved before or if it has, the mainstream literature does not reflect this. Second, I apply this to models with factorizing likelihoods, resulting in Sample Inflation, a technique to generate many dependend samples from few likelihood evaluations. The evaluation in section 6 showed that Sample Inflation can reduce variance and help to attain high likelihood regions more quickly in a Population Monte Carlo setting.
Future work will have to derive variance estimates for Sample Inflation and, as a consequence, measures of Effective Sample Size and perplexity \cite{Robert2010}. This will hopefully lead to a better understanding of when Sample Inflation can help and under which conditions it hurts performance.

\section*{Acknowledgments} 
I thank Ilja Klebanov for proof reading and very valuable suggestions with regard to notation and readability. Patrick Jähnichen and Shirin Riazy checked the proofs and provided helpful discussions, as did Christoph Teichmann. Christian Robert hinted at certain improvements in readability.


\bibliography{library}
\bibliographystyle{icml2015}

\end{document}